**Affective and cognitive prefrontal cortex projections to the lateral habenula in humans.**

Karin Vadovičová

Neuroradiology Unit, Department of Diagnostic Imaging, Spedali Civili, 25123 Brescia, Italy.

email: vadovick@tcd.ie

Abstract

 Anterior insula (AI) and dorsal ACC (dACC) are known to process information about pain, loss, adversities, bad, harmful or suboptimal choices and consequences that threaten survival or well-being. Also pregenual ACC (pgACC) is linked to loss and pain, being activated by sad thoughts and regrets. Lateral habenula (LHb) is stimulated by predicted and received pain, discomfort, aversive outcome, loss. Its chronic stimulation makes us feel worse/low and gradually stops us choosing and moving for the suboptimal or punished choices, by direct and indirect (via rostromedial tegmental nucleus RMTg) inhibition of DRN and VTA/SNc. The response selectivity of LHb neurons suggests their cortical input from affective and cognitive evaluative regions that make expectations about bad, unpleasant or suboptimal outcomes. Based on these facts we predicted direct dACC, pgACC and AI projections to LHb, which form part of an adversity processing circuit that learns to avoid bad outcomes by suppressing dopamine and serotonin signal. To test this connectivity I used Diffusion Tensor Imaging (DTI). I found dACC, pgACC, AI and caudolateral OFC projections to LHb. I predicted no corticohabenular projections from the reward processing regions: medial OFC (mOFC) and ventral ACC (vACC) because both respond most strongly to good, high valued stimuli and outcomes, inducing dopamine and serotonin release. This lack of LHb projections was confirmed for vACC and likely for mOFC. The surprising findings were the corticohabenular projections from the cognitive prefrontal cortex regions, known for flexible reasoning, planning and combining whatever information are relevant for reaching current goals. I propose that the prefrontohabenular projections provide a teaching signal for value-based choice behaviour, to learn to deselect, avoid or inhibit the potentially harmful, low valued or wrong choices, goals, strategies, predictions and ways of doing things, to prevent bad or suboptimal consequences.

## 1. Introduction

I examined the cortical input from the affective and cognitive prefrontal regions to the lateral habenula in humans. I predicted that dACC, pgACC and AI activate the LHb via direct and indirect projections, forming together an adversity processing circuit. This circuit biases learning, thoughts, feelings and behaviour towards gradual inhibition of harmful, punished or suboptimal choices by potentiating the D2 loop of ventral striatum (VS) and by suppressing dopamine and serotonin signal via LHb. Lack of dopamine strengthens the inhibitory avoidance learning and inhibitory self-control and weakens the motivation and drive to move and work for goals and rewards (Vadovičová and Gasparotti, 2013, **Fig. 1**). I suggest that overstimulation of LHb causes discomfort and aversion by down-regulating serotonin signalling (Wang and Aghajanian, 1977), lack of which disinhibits AI, dACC, GPi, LHb (presynaptic LHb inhibition found by Shabel et al., 2012) and pain pathway output, thus potentiates learned helplessness, depression and anxiety. This is supported by rat studies where learned helplessness was eliminated by habenular lesions, and correlated with increased LHb (and lateral septum) metabolic activity and synaptic potentiation (Amat et al., 1998; Li et al., 2011; Mirrione et al., 2014; Li et al., 2011).

I expected no corticohabenular projections from the reward processing regions, as the good, valuable, rewarding choices, appraised by mOFC, tend to move and motivate us to choose, act and go for them – by stimulating the motivational D1 loop of VS and dopaminergic VTA (**Fig. 1**). Similarly, when we are doing well, reaching good outcomes, safety and gains, it is signalled to brain by the vACC activation, that generates fulfilment and satisfaction, increases well-being and lightens up mood by inducing serotonin release in the brain (**Fig. 1**). Based on DRN afferents (Peyron et al., 1998; Vertes, 2004) and serotonin dependent antidepressant effect of vACC stimulation in rats (Hamani et al., 2010), I proposed that vACC stimulates DRN directly and via bed nucleus of stria terminalis (BNST). So the LHb is suppressed by dopamine and serotonin, by mOFC to VTA and vACC to DRN projections. In contrary, LHb stimulation and its activation by GPi input (Hong and Hikosaka, 2008) suppresses dopamine signaling in VTA/SNc (Christoph et al., 1986; Jhou et al., 2009; Hong et al., 2011; Matsumoto and Hikosaka, 2009) and serotonin in median and dorsal raphe nuclei (MRN, DRN), (Wang and Aghajanian, 1977), directly and via the inhibitory RMTg. The LHb suppresses also wake/arousal and locomotion promoting histamine release in rats, and opposes its effect on supramamillary nucleus (Kiss et al., 2002, Vanni-Mercier et al., 1984, Onodera et al., 1994). Lateral habenula projects also to noradrenergic locus coeruleus, reticular formation, lateral preoptic area, mediodorsal thalamus (MDT), lateral hypothalamus and superior colliculus (Herkenham and Nauta, 1979). The electrophysiology studies in macaques (Matsumoto and Hikosaka, 2009) showed that LHb neurons respond to punishment cues, unfavourable outcomes and reward omissions. They were most excited by the most negative of the available outcomes, thus firing inversely to the VTA/SNc neurons that are excited by expectation of valuable outcomes. So we claimed that while the reward processing circuit learns about good, valuable choices which increase our well-being and prospects, the adversity processing circuit learns about potentially bad, wrong, harmful or unpleasant choices and outcomes, which decrease well-being and survival chances. Good, interesting things/choices are linked to the approach and motivation to gain them, presumably by preferential mOFC input to D1 loop of VS, while bad things such as pain, harm and loss induce avoidance and aversion, by predominant AI, dACC and pgACC input to its D2 loop. So we extended the basal ganglia learning model (Frank and Hutchison, 2009) by adding the affectively biased cortical input to the VS.

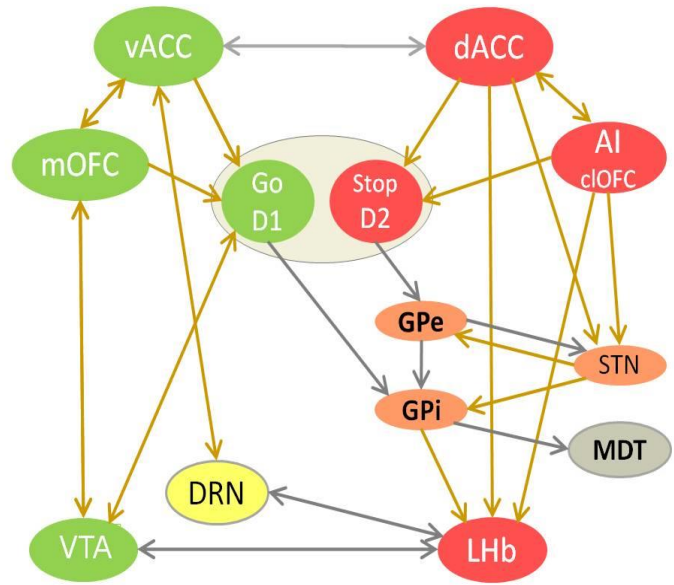

**Figure 1. The competition between reward and adversity processing circuit in choice behaviour.** This model of value-based learning shows how the affective processing causes selection of good/valuable and de-selection of bad/harmful choices, by controlling dopamine and serotonin signaling in the brain (Vadovičová and Gasparotti, 2013). The implicit bias/inclination for the 'Go for it' versus 'Stop yourself - avoid it' response is learned in the motivational ventral striatum (VS), through potentiation of cortical glutamatergic synapses by dopamine at D1 loop, and their depotentiation at D2 loop. Reward and adversity processing circuits are marked green and red. Neurons with either D1 or D2 receptors are spatially intermixed in VS. VS Neurons with D1receptors disinhibit dopamine neurons by inhibiting VTA GABA interneurons. The indirect D2 loop (in orange) biases the choice selection toward inhibitory avoidance. The prefrontal projections of mediodorsal thalamus (MDT) enforce the representations of choices and goals in working memory, as part of cortico-striato-thalamo-cortical loop. Dopamine source VTA is marked green, serotonin source DRN in yellow. The projections with excitatory effects are brown, with inhibitory effect are grey. Dopamine attenuates output of the adversity processing circuit (in red) and potentiates that of reward processing circuit (in green). Serotonin attenuates the AI, dACC, GPi, LHb, SNr, STN plus motivational D1 loop of VS and enforces vACC, SNc (via SNr inhibition), GPe.

This DTI study is based on the affective processing model that explained how the connectivity of the reward and adversity processing circuits causes their competition, their opposite effects on neuromodulators control, decision making, choice behaviour and well-being (Vadovičová and Gasparotti, 2013). This model combined wide evidence from functional, anatomical, dopaminergic, serotonergic and mental dysfunctions studies to specify the interaction of cortical regions with VTA, DRN, LHb, D1 and D2 loop of VS in value-based learning. It stated that dopamine signaling directs and drives us towards valuable, worthy – good, rewarding, interesting, novel, useful, important, relevant, right and meaningful things, choices and information. This model showed how dopamine guides us to choose, prefer, want, desire, engage with, get interested, inclined, even addicted (in love) to them, and to hope, seek, move, go and work for the valuable, survival or well-being promoting things (food, people, safety, affection, beauty, goals). It hypothesized that serotonin's role is to keep our consumption and wanting within the limits of homeostasis, and to signal when we reached the 'comfort zone'. Thus the optimal brain serotonin levels promote well-being, fulfilment, satisfaction, feeling well, alright, at ease, non-deprived. They also attenuate drive, motivation, impulsivity, motion and effort, calm down worries, aggression, pain, deprivation and slow us down to rest. The proposed vACC role in signalling gain, well-being, safety, achievement, success is supported by its activation by rewarding outcomes and fear extinction, plus its deactivation by melancholia (Grabenhorst and Rolls, 2011; Quirk et al., 2003; Pizzagalli et al., 2004).

I observed robust dACC and AI co-activation in our NoGo task and in the literature, towards things that are bad or harmful such as pain, loss, risk, suboptimal outcomes, rejection or distress (unpublished M.Sc. thesis). These aversive events decrease well-being and survival chances, so we learn to avoid them by de-selection of choices leading to bad consequences. This de-selection is done both consciously - by changing our goals and plans in the medial PFC, and unconsciously - by probabilistic learning of bad or wrong choices by basal ganglia, and by inhibition of dopamine and serotonin release via LHb – consequently affecting all brain regions with dopaminergic or serotonergic receptors. I claimed that evaluations and interpretations in the dACC and AI bias the response selection towards inhibitory avoidance by activating the LHb and D2 loop of VS. Dorsal ACC learns, predicts and warns us when we are not doing well, to prevent harm, loss (of resources, loved ones, time) or failure. Its warning signal induces worry, precaution and alarm state, leading to attention, alertness, mobilization (for fight or flight) in risky, speed or accuracy demanding situations. This warning signal from dACC urges the prefrontal cortex to switch away from the inadequate or faulty strategies, to think why things go wrong and find solutions how to change/adjust our world or behaviour to stop losing or getting harmed. The AI detects and reacts with aversion to bad, inferior or noxious quality of objects, subjects and social conduct, and also to their moral, conceptual, contextual or task related wrongness. For the pgACC, active in regret, sorrow and sadness (Drevets et al., 1992, George et al., 1995, Brody et al., 2001), I predicted similar LHb projections as for the dACC and AI, leading to passive avoidance.

To test the proposed corticohabenular projections in humans I used DTI probabilistic tractography. This method does not discriminate the afferent from efferent axonal fibres. But because the tracing studies in animals found only the corticohabenular, no habenulocortical projections, I assumed that the fibre tracts in this study are the LHb afferents. The medial PFC projections to LHb that regulate dopamine system were shown already in 1982 in rats (Greatrex and Phillipson, 1982). A tracing study in macaca fuscata found dACC but no vACC/BA 25 projections to LHb (Chiba et al., 2001). Fronto-habenular projections were also shown by DTI tractography in humans (Shelton et. al, 2012), seemingly originating in BA 10 in their Figure. A retrograde and anterograde tracing study (Kim and Lee, 2012) in rats found corticohabenular projections from the AI, cingulate, prelimbic and infralimbic cortex. The authors found that dense descending projections terminating in the MDT made *en passant* and terminal projections to the LHb. The PFC is reciprocally connected with MDT, forming cognitive and affective cortico-thalamo-cortical loops passing via dorsal (DS) and ventral striatum (Alexander et al., 1986). The infralimbic cortex in rats is a homologue of vACC in humans, while prelimbic cortex is a homologue of dACC.

2. Materials and Methods

I used 3 Tesla DTI datasets of 18 healthy participants (24-30 years old) obtained from the NKI Rockland Sample as part of the 1000 Functional Connectomes Project (http://fcon_1000.projects.nitrc.org/indi/pro/nki.html). The DTI data were acquired with 137 gradient directions, 2mm isotropic voxels, 64 slices and FOV 106 x 90 mm. The T1 weighted anatomical images were acquired with TR/TE/TI = 2500/3.5/1200ms, FOV 256 x 256 mm, flip angle 8 degrees and 1 mm isotropic voxels.

For DTI analysis was used the FSL (FMRIB Software Library, http://www.fmrib.ox.ac.uk/fsl/) version 4.1.9 (Jenkinson et al., 2012; Smith et al., 2004; Woolrich et al., 2009), with a Probtrackx tool (Behrens et al., 2007) for probabilistic tractography. This method generated probabilistic connectivity distributions for the tested axonal projections for each participant. Each DTI dataset has been analyzed independently, using standard FSL procedures. The pre-processing steps included brain extraction, head motion (Jenkinson and Smith, 2001; Jenkinson et al., 2002) and eddy current correction (Behrens et al., 2003). Then Bedpostx tool was applied to calculate diffusion tensor and to model crossing fibres within each voxel of the brain. Default statistical threshold was used for all analyses. The results were coregistered to the anatomical image and then normalized to the FSL MNI template (MNI152 2mm). The same procedure was repeated on 3 additional 1.5 Tesla datasets. The axonal connectivity was analyzed in individual brains instead of in group, to avoid smoothing, which would confound the habenular voxels with adjacent regions.

The seed and target regions for probabilistic tractography were selected manually in the right hemisphere of each brain using the anatomical image (Fig. 2). Right hemisphere was chosen as sufficient for my research question, plus some fMRI studies reported its stronger aversive response (Simon-Thomas et al., 2005). The main seed regions of interest were in the AI, dACC and pgACC. The target region was in the LHb, discriminable by its contrast difference due to myelinated fibres. The reward processing mOFC and vACC regions were also tested, as I hypothesized their lack of input to LHb. The vACC seed region contained the Brodmann area (BA) 25, which is the posterior part of vACC, plus few voxels adjacent to it but posterior to pgACC. The mOFC seed contained the posterior half of the gyrus rectus, to avoid the adjacent ventral BA 10. The exploratory seed regions sampled the remaining prefrontal cortex regions: BA 8, 9, 10, 12, 44, 45, 46, 47, to test the input from cognitive prefrontal areas to the LHb. When drawing the LHb seed regions I avoided the voxels adjacent to the ventricle, but included remaining MHb voxels, as the MHb/LHb border is not discriminable in the anatomical image. This inclusion limits but should not affect my prefrontohabenular connectivity results, as tracing studies found no PFC-MHb connectivity in rats (Herkenham and Nauta, 1977).

### 3. Results

My probabilistic tractography results confirmed the predicted projections from AI, dACC and pgACC to the LHb in all 18 participants (Fig. 3 to 9). Strong interconnectivity was observed between the AI and the adjacent caudolateral OFC (clOFC). Based on shared connectivity, they seemed more similar to each other than to other regions. This was found in all 18 studied brains and suggests their common role in inhibitory avoidance. The lack of corticohabenular projections from the reward processing regions was confirmed in all 18 participants (Fig. 10) for vACC and in 15 for mOFC. Questionable, probably disynaptic fibre tracts between mOFC, BA 10 and LHb were found in remaining 3 brains. Because the ventral BA 10 to LHb tract was adjacent to the reciprocal BA10/mOFC connections, I could not discriminate these tracts in 3 brains, for which the results were inconclusive. The indirect, disynaptic tracts in DTI results are caused by misidentification of the fibre tracts crossing same voxel. I found clear disynaptic fibre tract between the mOFC and hypothalamus, and hypothalamus and LHb. My prefrontohabenular connectivity findings were repeated with 1.5 T datasets, using same probabilistic tractography analysis in 3 additional participants (Fig. 11 to 14).

The prefrontal fiber tracts to the lateral habenula passed mostly via the internal capsule, some via basal ganglia and all via ventromedial part of the anterior thalamus (AM). It is possible that PFC-LHb projections formed collaterals in AM, as they showed second highest fibre tract density there, after the LHb. The same projections were found after applying the exclusion masks to exclude the MDT and superior colliculus, located above and under the LHb, from the analysis. The dorso-ventral position of the cortical tracts on their way through capsula interna depended on the vertical position of the individual cortical seed regions. So, the ventral BA 10 projections to LHb crossed striatum between the nucleus accumbens and ventral anterior putamen, forming horizontal fiber tract. But projections from more dorsal regions such as dACC or BA 9 crossed striatum between the lateral caudate nucleus and putamen, forming diagonal tract. The temporal pole and an additional fibre tract from AI reached the LHb not via capsula interna but by posteriorly localized tracts (Fig. 15, 3).

Unexpected findings of this study were the strong projections to LHb from the cognitive PFC regions: from seeds in the superior, middle and inferior frontal gyrus and from the medial and lateral frontal pole or BA 10. These prefrontohabenular fiber tracts projected from the BA 10, 9, 8, 44, 45, 46, 47, FEF and lateral OFC (Fig. 12 and 13). I found corticohabenular fibre tracks from all cognitive and affective (AI, clOFC, dACC, pgACC) PFC regions involved in decision making except the vACC and probably except mOFC.

I showed direct projection between septum and medial habenula MHb (Fig. 14) passing via bed nucleus of stria terminalis (BNST) and AM and possibly branching there. The MHb formed the known multisynaptic fibre tract with the hippocampus: hippocampus → fornix → septum → Hb → pineal gland. I found also hypothalamo-LHb fibre tracts in humans, without examining exact nuclei. Similar connectivity in rats includes LHb afferents from the lateral hypothalamic and preoptic nucleus, plus efferents to the lateral preoptic and supramamillary nucleus, ventrolateral septum, lateral and dorsomedial hypothalamus (Herkenham and Nauta, 1977, 1979).

The following figures show a sample of found fibre tracts. These DTI images are from different individual brains. The left hemisphere in the image represents the right hemisphere of brain. The LHb slice is not always shown, as I prioritized to show fibre tract passage.

**Fig. 2. The localisation of my ACC seed regions and habenula. (A)** The cross marks the habenula position in the spm anatomical template colin27 (Evans et al., 1993)**.** The coronal section passes through the posterior commissure. **(B)** The approximate positions of seed regions are shown in sagittal image of human brain from Ongur et al., 2003. In red is dACC seed region (dorsoposterior to genu of corpus callosum), in pink pgACC (anterior to genu), in yellow vACC (ventroposerior to genu). **(C)** Right and left habenula has lighter contrast than its surrounding. **(D)** The LHb region targeted by cortical fibre tract.

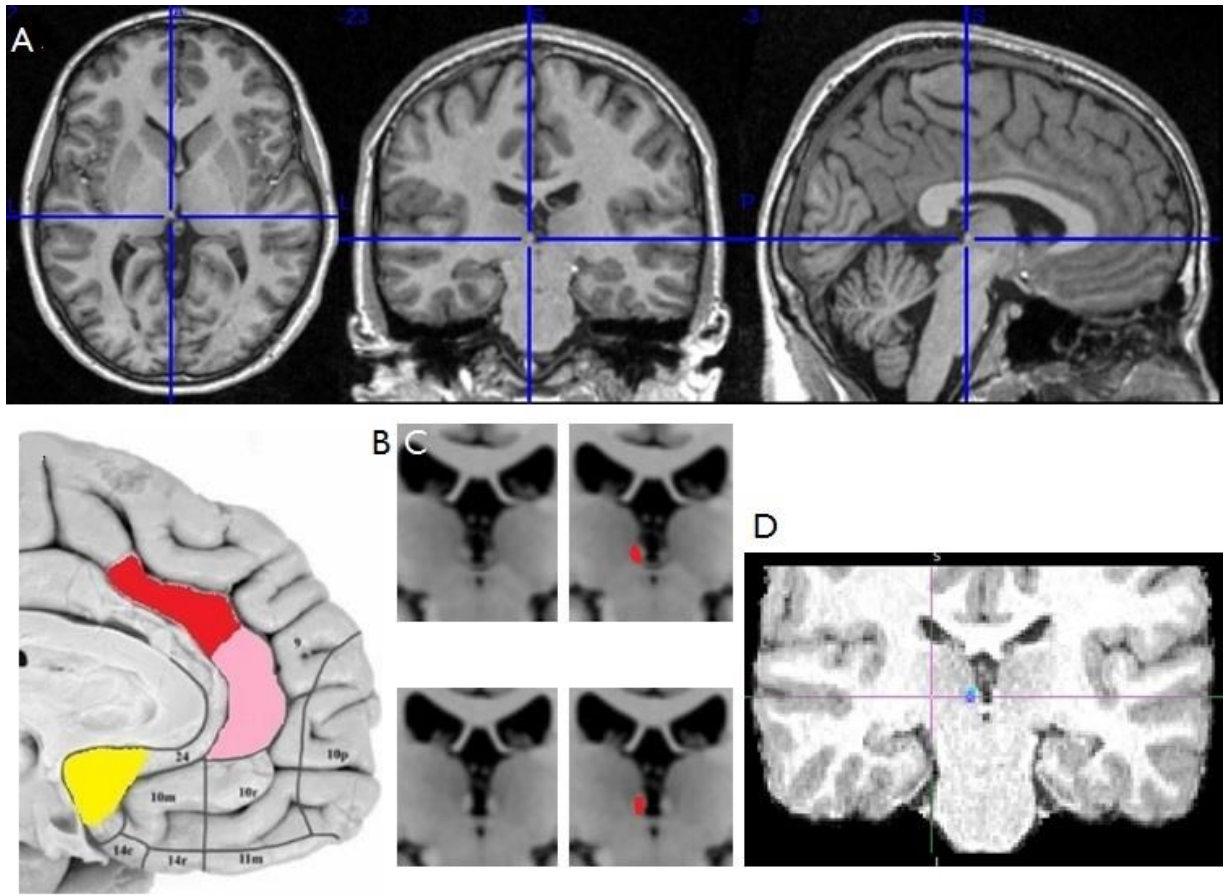

**Fig. 3. AI to LHb fibre tract. (A, B)** The AI tract is branching to clOFC that is anterior to ventral AI. The arrows point to AI. **(C)** Visible is branching from AI to temporal pole (TP) and TP to LHb.

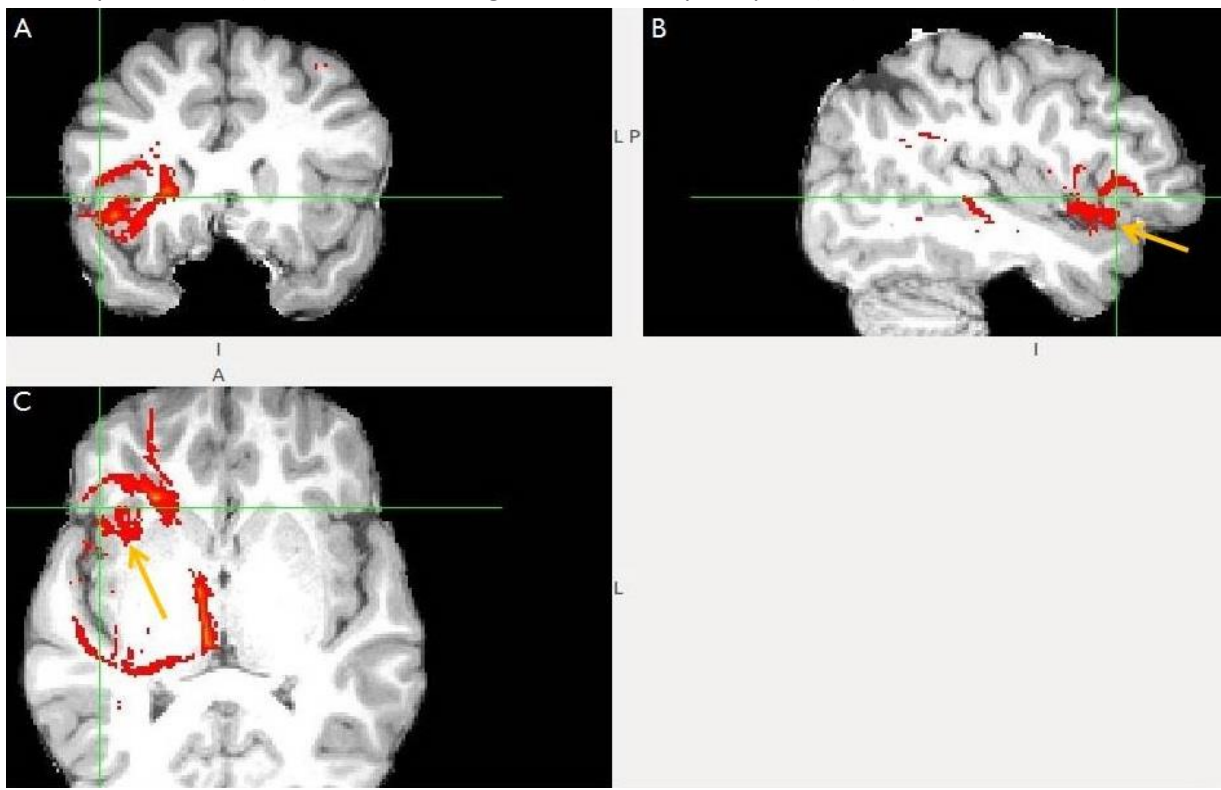

**Fig. 4. AI to LHb fibre tract, different slices. (A)** Tract termination in right and left habenula (arrow). **(B)** AI (arrow) is connected with clOFC and projects to temporal pole. **(C)** AI/clOFC to LHb.

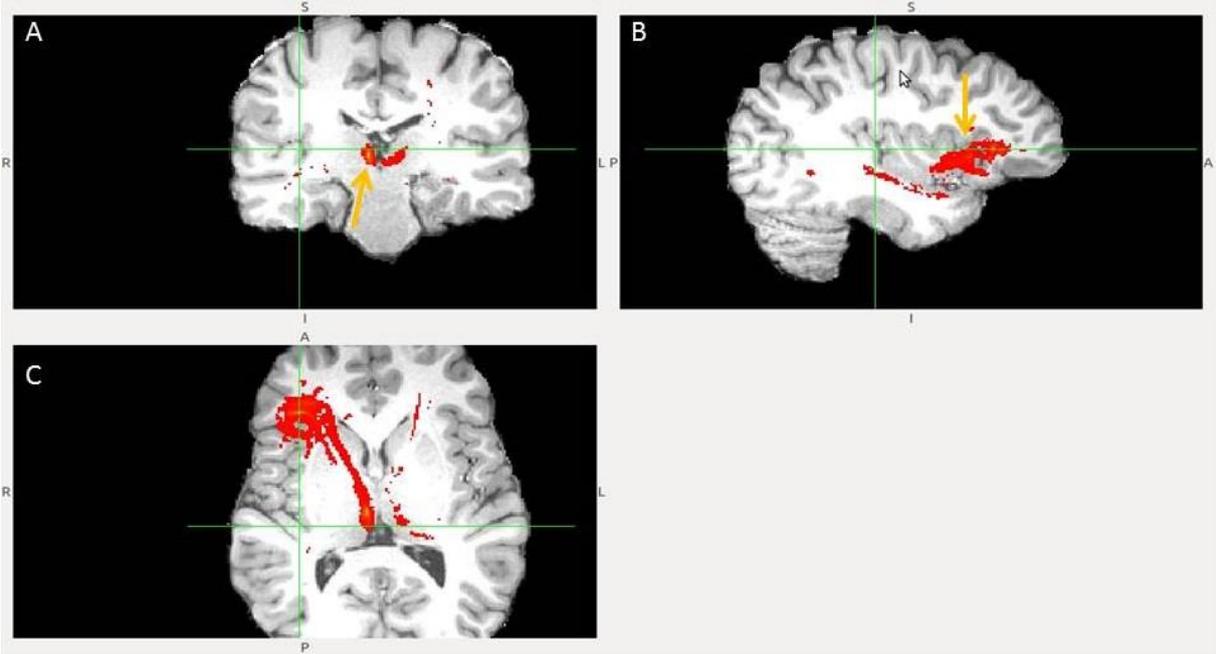

**Fig. 5. AI to LHb fibre tract, anoher slices. (A)** AI and clOFC projections to LHb via capsula interna. **(B)** AI tract decends to clOFC. (C) AI/clOFC tract targets LHb, shown at its dorsal part.

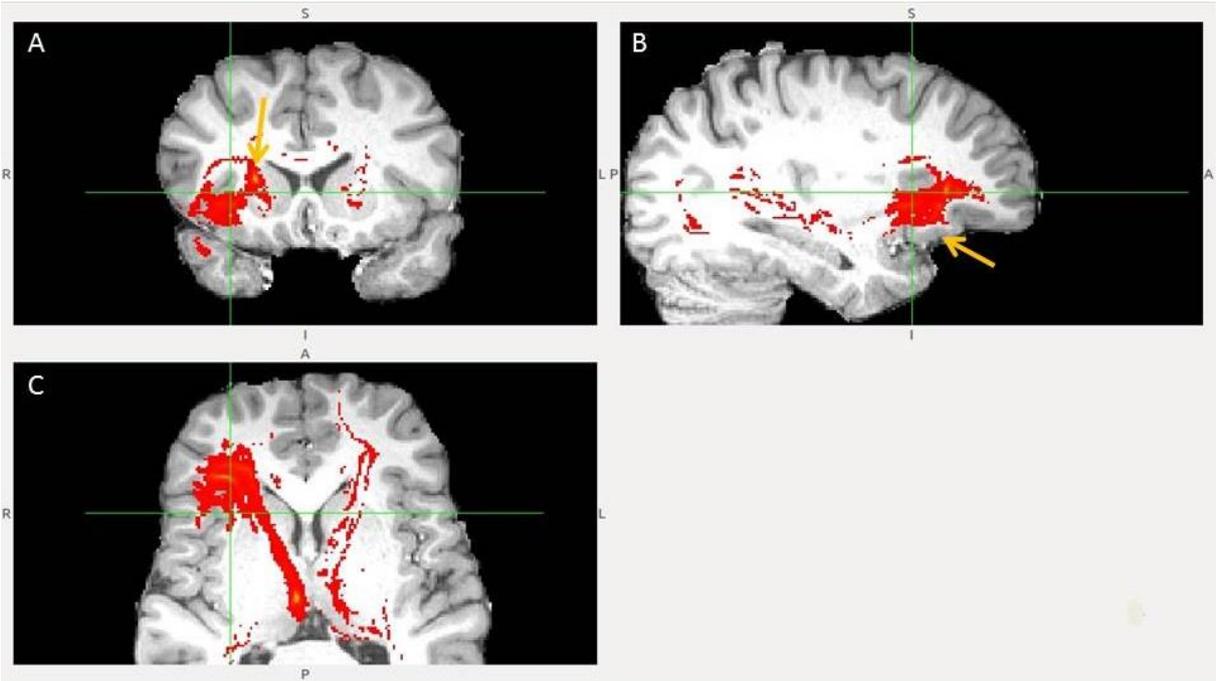

**Fig. 6. Dorsal ACC to LHb fibre tract. (A)** Tract passes via capsula interna. **(A, B, C)** The images show dACC, pointed to by arrow.

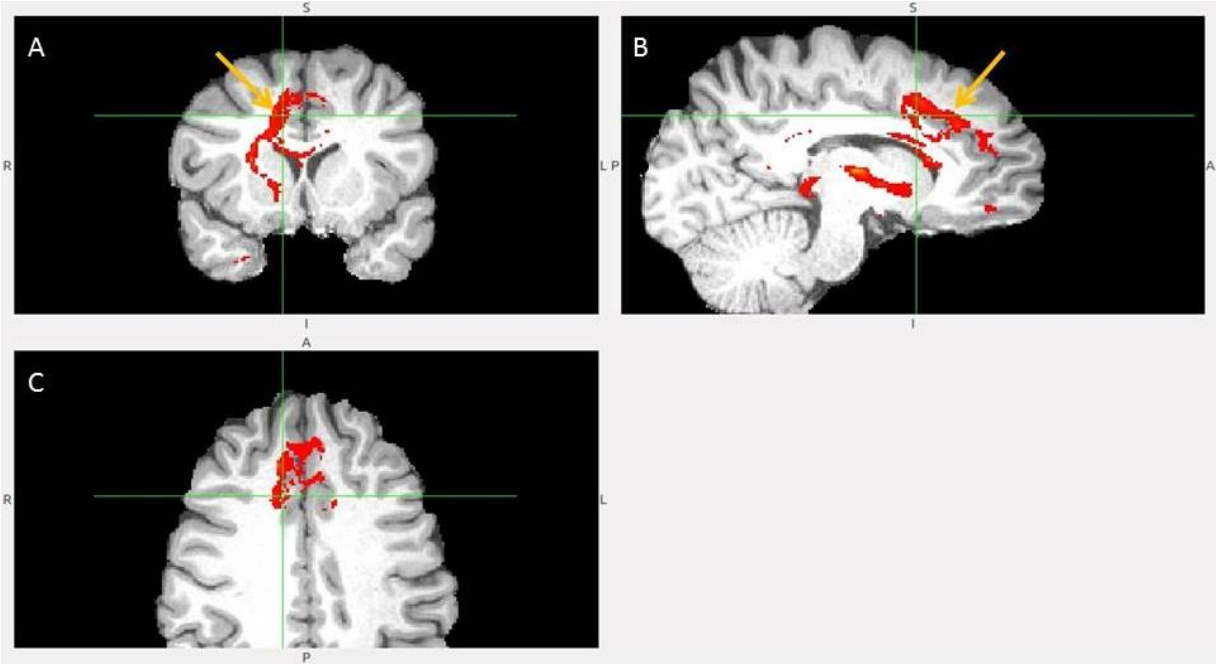

**Fig. 7. dACC to LHb fibre tract, different slices.** Arrows indicate dACC **(A, B)** and LHb **(C)**.

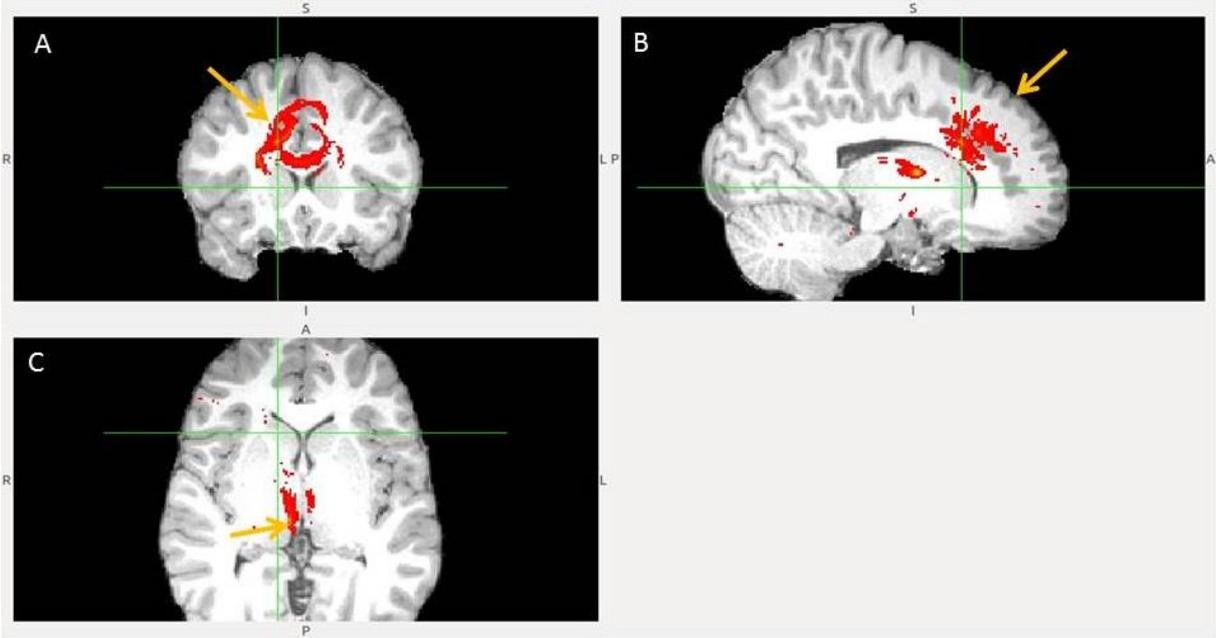

**Fig. 8. pgACC fibre tract to LHb. (A)** pgACC pointed by arrow. **(B)** pgACC is located in front of genu of corpus callosum. This fibre tract passes via medial anterior thalamus (AM) to LHb, then branches to superior colliculus, PAG, DRN, pons. **(C)** Arrows indicate ventral part of pgACC and part of tract just above the LHb.

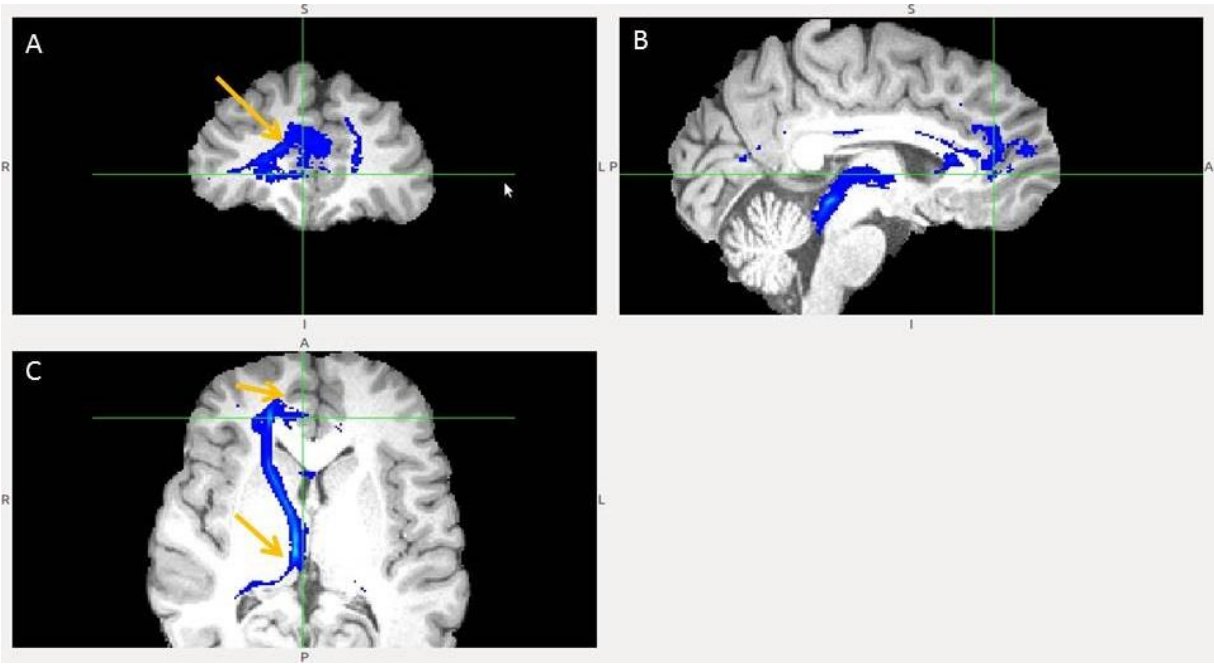

**Fig. 9. pgACC fibre tract to LHb, another slices. (A)** Frontal pole cluster where pgACC tract is branching. **(B)** Arrow points to pgACC part of the tract. Visible is thalamic passage. **(C)** Arrows show pgACC and capsula interna part of tract.

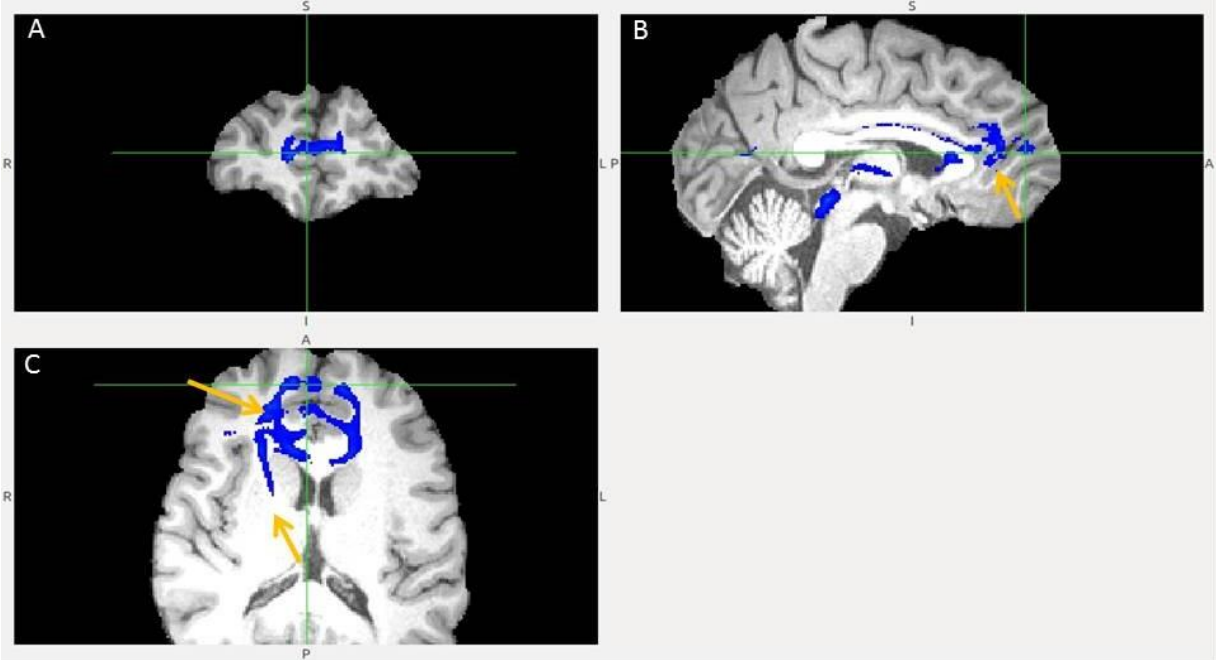

**Fig. 10. No vACC to LHb fibre tract.** This tract **(B, C)** reached LHb only indirectly by its projections to BA 10 **(A)**. BA 10 has own robust LHb projections. Ventral ACC is indicated by lower arrow, BA 10 by upper one **(C)**. Ventral ACC tract branched also to hypothalamus, which has reciprocal connection with LHb (not shown).

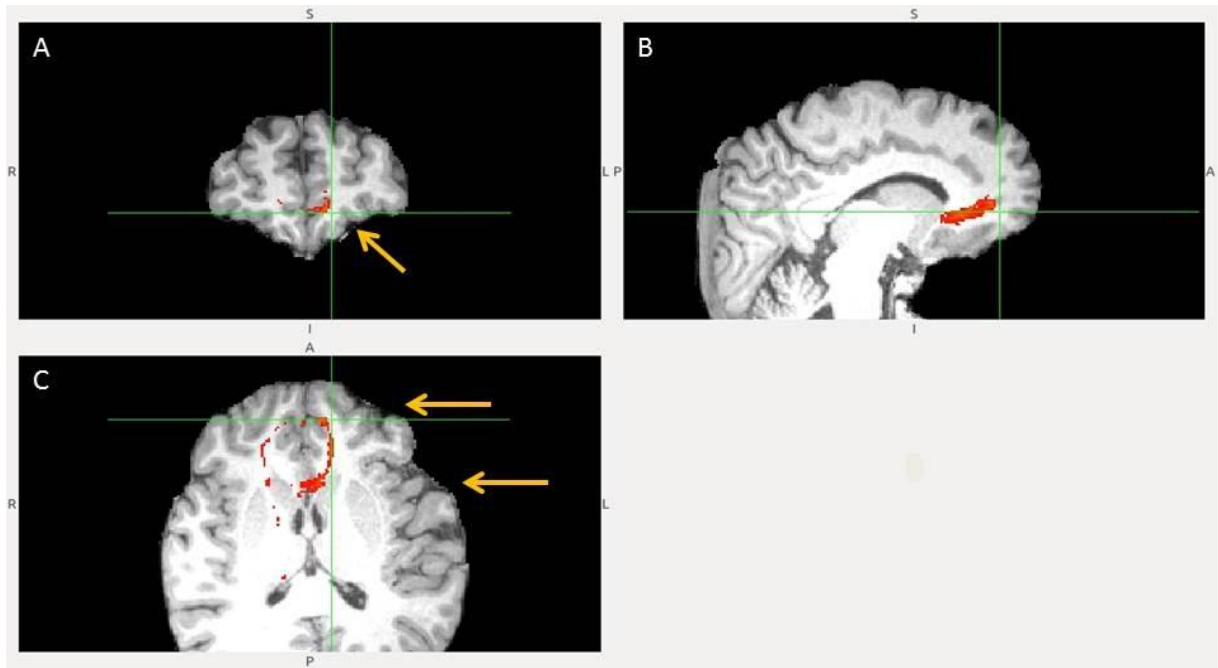

**Fig. 11. pgACC to LHb tract.** The fibre tract is in red, pgACC seed region in orange. **(A)** Passage via capsula interna. **(B)** Upper arrow points to pgACC seed, bottom left to LHb and bottom right to tract passing via AM. Visible is also the pgACC projection to dACC, which projects to LHb on its own. **(C)** Arrow shows where the tract descends into LHb, just 1 slice above it.

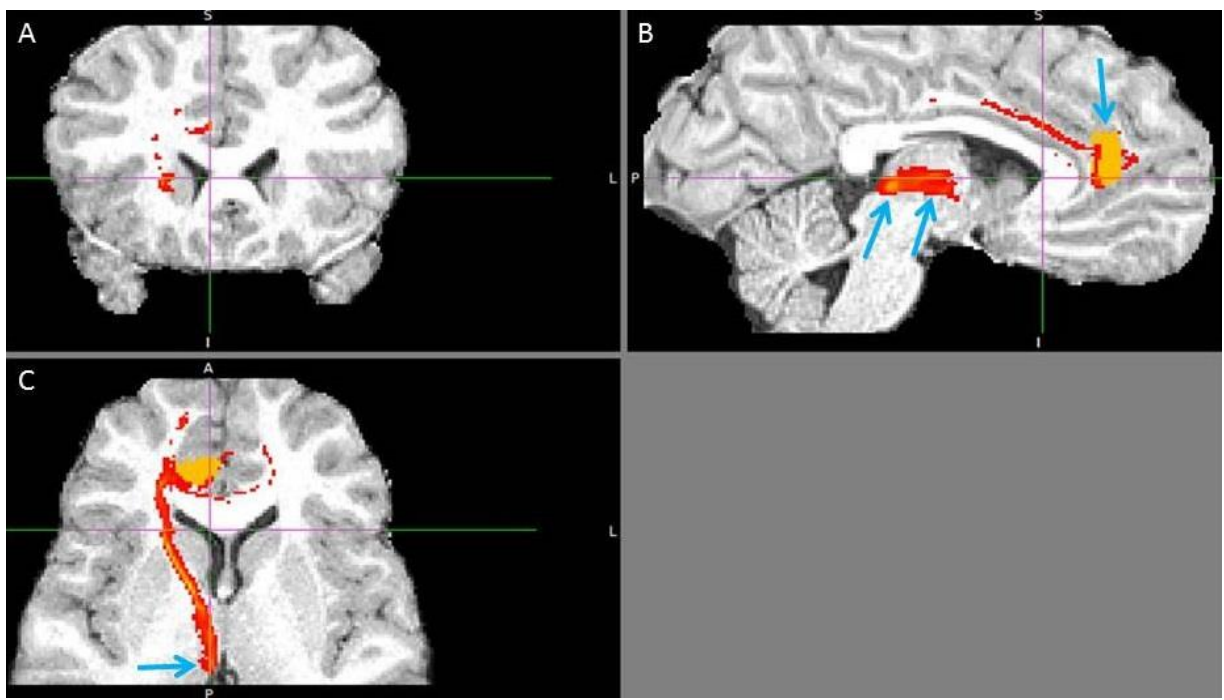

**Fig. 12. Frontal pole to LHb projection.** Arrows point to LHb **(A)** and to tract just above it **(B).**

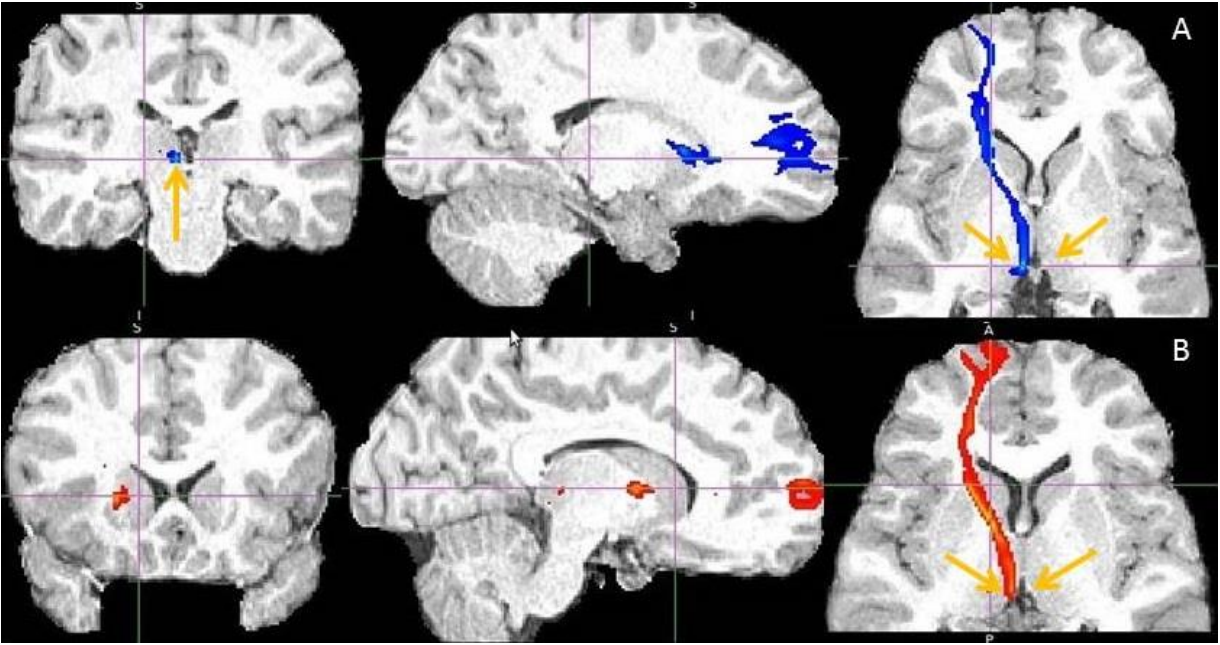

**Fig. 13. Lateral BA 10 (red) and inferior frontal gyrus IFG (blue) fibre tracts to LHb. (A)** Arrow indicates LHb. **(B)** The highest density of connected fibres is in LHb (purple region). Both seed regions have overlapping tracts via thalamus (right arrow shows AM). **(C)** Arrow points above LHb.

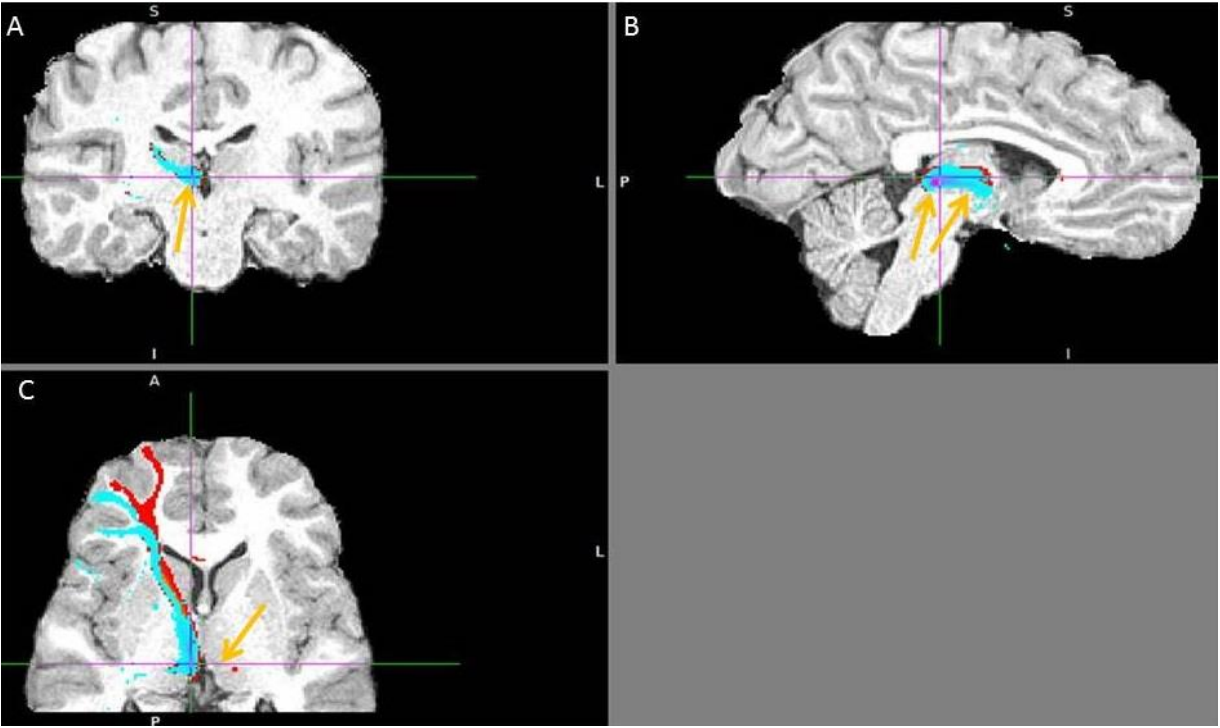

**Fig. 14. hippocampus to medial habenula (MHb) fibre tract.** This fibre tract includes:

hippocampus → fornix → septum → MHb → pineal gland. The tract between septum and habenula passes via AM and via BNST, where it might send axon collaterals. The AM has the second highest fibre density in this tract. The BNST might connect both septum and habenula via afferents, efferents or both. Fornix has known projections to septum, anterior thalamus and mamillary body, and septum projects to MHb. **(A, D)** Tract passage via septum and AM. Arrow marks amygdala, linked by tract with hippocampus **(B, E, H)** Tract passage via AM to MHb, and MHb to pineal gland. Arrow marks MHb. **(C, I)** Tract passage via fornix to septum, and septum via AM to MHb. Arrow in **(C)** points to AM, in **(I)** arrow points to fornix and MHb. **(F)** tract passage via hippocampus, towards fornix posteriorly and amygdala anteriorly. **(G)** The arrows point to septum (upper) and MHb (lower).

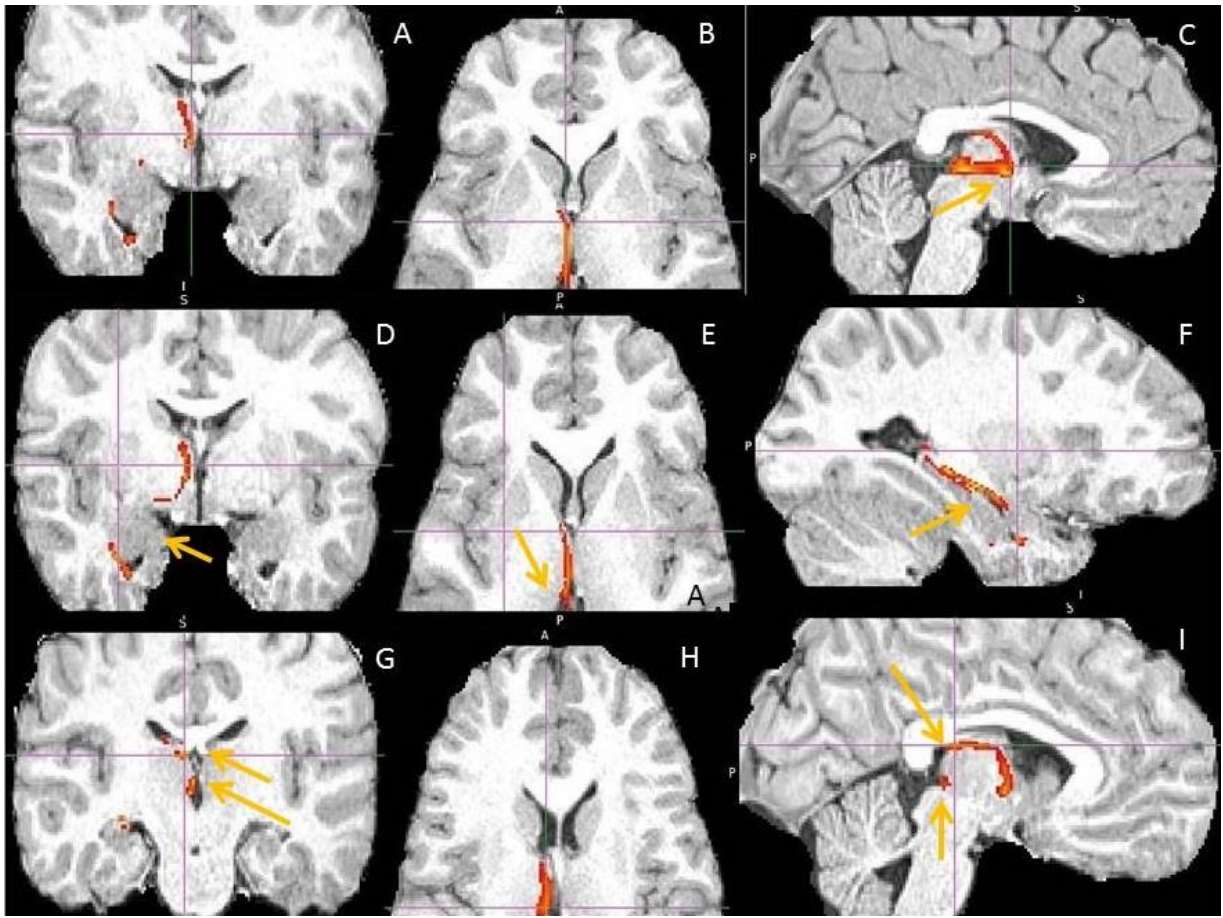

**Fig. 15. temporal pole to LHb fibre tract. (A)** The arrows point to habenulae. **(B)** Tract passage from temporal pole (TP) via temporal lobe. **(C)** The arrows mark TP and its tract towards superior colliculus and up to LHb.

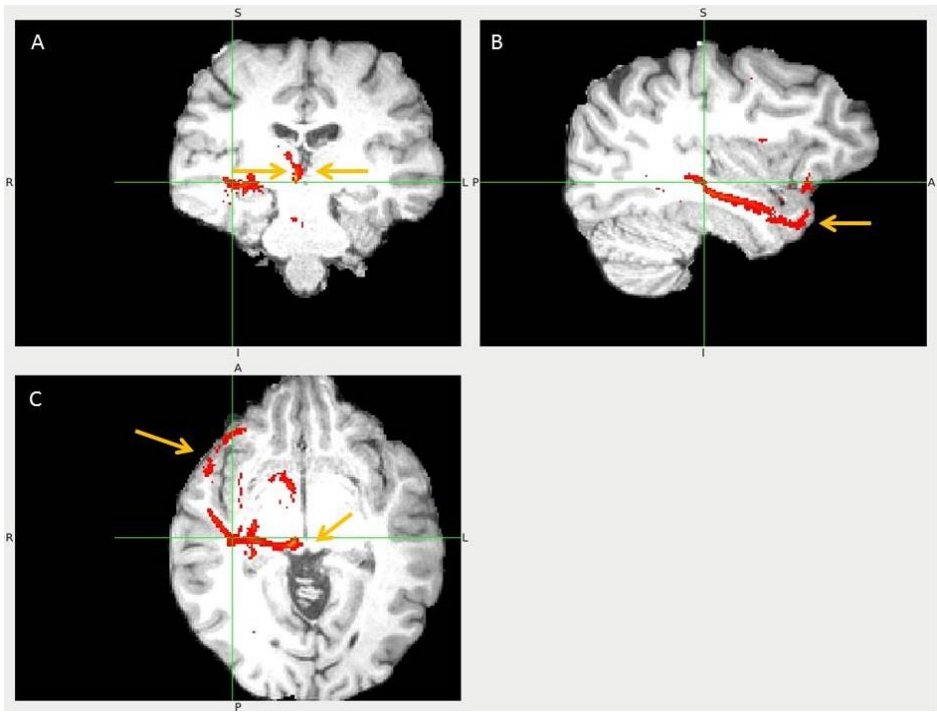

### 4. Discussion

My DTI tractography results support the functional connectivity of proposed adversity processing circuit, formed by dACC, pgACC, AI and adjacent clOFC input to the LHb. This circuit detects, learns about and predicts potential adversities and forwards the information about bad, harmful or suboptimal choices and consequences to the LHb, to suppress dopamine and serotonin release in the VTA and DRN (Vadovičová and Gasparotti, 2013). The causal role of this corticohabenular circuit in aversive processing is supported by findings from functional, behavioural, pharmacological and mental disorders studies. An example is the co-activation of dACC, AI and LHb during negative feedback (Ullsperger and von Cramon, 2003) in humans.

I observed strong interconnectivity between the AI and adjacent caudolateral OFC (clOFC). Based on their shared connectivity and common co-activation in functional tasks in literature, I suggest that the AI and clOFC form a functional processing module, because connectivity determines function and computational role. Possibly they process similar kind of information about bad - harmful, aversive, inferior, corrupted or suboptimal qualities/attributes of objects, subjects or conducts. So the AI/clOFC probably selectively learns about things of negative or low value - aversive, unpleasant, disliked or safer to avoid. The AI/clOFC responds also to the contextual (regarding current task or situation), conceptual (false, strange, misfit) and moral wrongness of things or conduct.

In accordance with the affective circuit competition model (Vadovičová and Gasparotti, 2013), in which the mOFC and vACC suppress the LHb via VTA and DRN stimulation, I found no direct vACC and likely no direct mOFC projections to LHb. The mOFC fibre tracts were either passing to LHb via hypothalamus (clearly not a direct projection) or via the ventral BA 10. The lateral hypothalamus is reciprocally connected with LHb (Herkenham and Nauta, 1977) and the mOFC with hypothalamus and BA 10. The anterior half of gyrus rectus, anteriorly adjacent to mOFC region, belongs to the ventral BA 10. The ventral BA 10 has strong projections to the LHb, so the questionable mOFC to LHb fibre tracts found in 3 of 18 participants were possibly formed by the reciprocal mOFC/BA 10 projections, adjacent to the separate BA 10 to LHb projections. This study supports the opposite effects of the adversity versus reward processing circuit on the activation versus inhibition of LHb. Evidence for the reward value coding in mOFC comes from many studies where mOFC responded to rewarding or pleasant stimuli and wins while lateral OFC responded to aversive options, punishment or loss (O'Doherty et al., 2001).

Temporal pole projections to LHb found in this study are likely from the neuronal populations linked to the negative meaning of the represented stimuli, attributes and concepts. Temporal poles process conceptual and semantic information about meanings, attributes and identities of things and persons. Temporal poles are interconnected with BA 10, dACC, vACC, mOFC, AI, thus receive information about both bad and good values and outcomes linked to objects, subjects or concepts.

I found a multisynaptic fibre tract passing from hippocampus via fornix to septum and from septum to MHb. Posterior septum is the main MHb afferent (Herkenham and Nauta, 1977). Septum was connected also with vACC and hypothalamic nuclei in my study. This septo-habenular fibre tract passed via AM and BNST. So septum might also be linked with BNST and BNST with habenula, by afferents, efferents or both. Hippocampus has known efferents from subiculum via fornix to septum, anterior thalamus and mamillary body (MB), plus MB also projects to AM (Aggleton et al., 2005). Supramamillary nucleus (SUM), which projects to medial septum and via fornix to hippocampus, receives afferents from LHb, interpeduncular nucleus (IPn), median raphe (MRN), DRN, preoptic areas (I guess IPn inhibits SUM), laterodorsal tegmentum (LDT), medial and lateral septum (Kiss et al., 2002). SUM is known to stimulate theta rhythm during exploration in rats (Vertes and Kocsis, 1997) that gets disrupted by MRN serotonin. The MHb receives input from glutamatergic, cholinergic or substance P neurons of the triangular septal and septofimbral nucleus, from GABAergic medial septum and diagonal band nucleus, dopaminergic VTA, serotonergic raphe and noradrenergic locus coeruleus neurons (Herkenham and Nauta, 1977, Qin and Luo, 2009, Gottesfeld, 1983). Medial habenula projects to the LHb (Kim and Chang, 2005), pineal body (Ronnekleiv and Moller, 1979), IPn and via IPn to MRN and LDT in rats (Herkenham and Nauta, 1979; Groenewegen et al., 1986). The MHb pathway regulates sleep cycle (Hikosaka, 2010). Based on its anatomical connectivity and interactions with the neuromodulators I propose that MHb stimulates the non-REM sleep via interpeduncular (IPn) and median raphe (MRN) nucleus, suppressing the REM, theta and alertness driving regions. This is supported by dense mu opioid receptors (which bind morphine) and circadian rhythmicity of MHb neurons (Guilding and Piggins, 2007; McCormick and Prince, 1987; Quick et al., 1999), by the markedly increased MHb or LHb activity during anesthaesia (Herkenham, 1981; van Nieuwenhuijzen et al., 2009; Abulafia et al., 2009), by the fact that the MHb neurons produce melatonin (Yu et al., 2002) and sleep promoting interleukin IL-18 (Sugama et al., 2002) and control via IPn the median raphe serotonin (Agetsuma et al., 2010; Wang and Aghajanian, 1977). Further evidence comes from high firing rates of MRN cells in non-exploratory waking states (when not recording new information in hippocampus) and in slow-wave sleep (SWS), plus their low firing rates during the exploration and in rapid eye movement (REM) sleep (theta states), (Jacobs and Azmitia, 1992; Marrosu et al., 1996). Serotonergic DRN neurons, noradrenaline and histamine neurons fire most during wakefulness, less during SWS and are suppressed in REM sleep (Hobson et al., 1975, McCarley et al., 1975; Pace-Schott et al., 2002), while cholinergic activity in LDT and pedunculopontine tegmental nuclei (PPT) is high at wake and REM sleep.

The unexpected findings of this study were the robust projections to LHb from the cognitive PFC regions, known for flexible coding, combining and holding in working memory any kind of relevant information useful for current goals and tasks (Rainer et al., 1998; Asaad et al., 2000). The strongest prefrontohabenular fibre tracts were from the frontal pole, also known as BA 10. Brodmann area 10 is interconnected with all PFC regions plus with the associative cortex in the temporal poles and superior temporal gyrus – so well informed and suited for flexible learning, reasoning, planning and goal-directed control of behaviour. The medial BA 10 is interconnected with the hippocampus and linked to temporal context, introspection, intentions, goals and planning – so to hierarchical temporal organization of our thoughts. The lateral BA 10, stimulated by informational novelty and problem solving, is most extended in humans. It probably induces dopamine release in the medial SNc that stimulates novelty seeking via motivational D1 loop of VS and goal-pursuit via D1 loop of medial head of caudate. I propose that lateral BA 10 role is to seek and find out what is going on - contingencies, patterns, rules, links between causes and consequences - to make cognitive predictions about our world, to test them and apply the right guesses/hypotheses and ways of doing things to reach our goals. Dorsal PFC is linked to spatial context, spatial organization, planning and control of behaviour, while the ventrolateral PFC is guiding thoughts and behaviour using meanings, ideas and interrelations between things, actions, events.

The robust input from the cognitive PFC to the LHb possibly enables the inhibitory self-control and context/goal dependent de-selection of wrong, irrelevant or inappropriate information, ideas, decisions, plans, strategies or ways of doing things, depending on the current task or goal. All prefrontal regions generate predictions about the world. The affective regions predict the reward value of choices and consequences, to bias decision making and selection of goals/intentions in the medial BA 10. These goals are then used by all PFC regions to plan and guide the execution of behaviour to reach goals and avoid harm or loss. The cognitive regions predict the informational value and cognitive significance of things and events - what is the optimal option or solution, right or wrong, useful, interesting, what has a predictive value.

Based on found connectivity and wide literature data I propose that affective prefrontohabenular input inhibits VTA, leading to potentiation of the D2 loop of VS, causing inhibition, passive avoidance and de-selection of harmful or suboptimal choices. Similarly, the cognitive prefrontohabenular input inhibits SNc, serving as teaching signal that potentiates the D2 loop of caudate head, to gather evidence on what went wrong, failed or was incorrect, to bias decisions via cortico-striato-thalamo-cortical loop. So the right, correct, valid predictions, ideas, models, strategies and ways of doing things (to reach goals), the "know how", IF-THEN rules or algorithms (manuals) are learned by the D1 loop of caudate head potentiation by increased dopamine, after the evidence proved them right/correct. The evidence for wrongness of the same prediction, hypothesis or the way of solving task is inferred from the strength of the glutamatergic synapses on the D2 loop of caudate head, which summate the experienced negative outcomes or failures, errors (for example in grammar rules learning in humans or in instrumental learning in animals). So the evidence "for" versus "against" the validity/correctness of the current prediction, guess, strategy, model or rule is memorized by the D1 versus D2 loop strength, to learn the probabilities of being right or wrong, or doing things the right or wrong way (for given context, situation).

In addition, the affective and cognitive prefrontal input to LHb possibly causes de-selection of the non-valuable choices and information from the working memory by suppressing dopamine release in the PFC. So the LHb activation caused by PFC input might decrease the dopamine signal (from SNc) in the PFC towards boring or irrelevant, non-significant information, deselecting the representation of uninteresting information in working memory, to occupy our attention with more useful information - depending on subjective values, motivations, priorities, intentions and goals. The useful, relevant, meaningful information induce dopamine release from SNc, to direct us toward interesting and informationally valuable stuff (i.e. information of predictive value). In accordance with this, the SNc projects to the cognitive PFC and dorsal striatum (Porrino and Goldman-Rakic, 1982; Haber et al., 2000) and the TMS stimulation of dorsolateral PFC (Strafella et al., 2001) induces dopamine release in the caudate nucleus.

Finding direct cognitive PFC projections to LHb means there might be some prefrontal neuronal populations that preferentially project to the LHb and other that project to SNc, depending on the value, meaning, significance or usefulness of the processed information for current goal or task. By their SNc and LHb efferents, the cognitive PFC regions bias our learning, selection and de-selection of information depending on their meaning and predictive value for current goal/aim, task or context.

## 5. Conclusions

Using DTI probabilistic tractography I found cortical projections to LHb in humans from affective PFC regions: AI, clOFC, lOFC, dACC and pgACC, linked to inhibitory self-control and inhibitory avoidance learning. As predicted I found no LHb projection from vACC. Unexpected findings were the robust projections to the LHb from temporal pole and from the cognitive prefrontal regions: BA 10 medial and lateral, BA 9, 8, 44, 45, 46 and 47.